\documentclass[nofootinbib]{revtex4-2}

\usepackage{scrextend}
\usepackage{graphicx}
\usepackage{indentfirst}
\usepackage{latexsym} 
\usepackage{multirow}
\usepackage{epsfig}
\usepackage{hyperref}
\usepackage{color}
\usepackage[usenames,dvipsnames]{xcolor}
\usepackage{amssymb}
\usepackage{amsfonts}
\usepackage{amsmath}
\usepackage{bm}
\usepackage{mathrsfs}
\usepackage{cleveref}

\usepackage[normalem]{ulem}
\useunder{\uline}{\ul}{}

\makeatletter
\newcommand{\multiline}[1]{%
  \begin{tabularx}{\dimexpr\linewidth-\ALG@thistlm}[t]{@{}X@{}}
    #1
  \end{tabularx}
}
\newcommand{\bea}{\begin{eqnarray}}
\newcommand{\eea}{\end{eqnarray}}
\makeatother

\def\cP{ {\cal P} }

\newcommand{\TSrb}{TS$_{\text{RB}}$}
\newcommand{\TSreg}{TS$_{\text{reg}}$}

\newcommand{\hLS}{$h_{\text{LS}}$}
\newcommand{\hNR}{$h_{\text{NR}}$}
\newcommand{\hML}{$h_{\text{ML}}$}

\newcommand{\cI}{{\cal I}}
\newcommand{\be}{\begin{equation}}
\newcommand{\ee}{\end{equation}}
\newcommand{\blank}{\bigskip}

\newcommand{\bigo}[1]{{\cal O}({#1})}

\def\addFaMAF{Facultad de Matem\'atica, Astronom\'ia, F\'isica y Computaci\'on,\\
 Universidad Nacional de C\'ordoba, (5000), C\'ordoba, Argentina}

\begin{document}

\title{Gravitational wave surrogates through automated machine learning}

\author{Dami\'an Barsotti}
\affiliation{\addFaMAF}

\author{Franco Cerino}
\affiliation{\addFaMAF}

\author{Manuel Tiglio}
\affiliation{\addFaMAF}

\author{Aar\'on Villanueva}
\affiliation{\addFaMAF}

\begin{abstract}
We analyze a prospect for predicting gravitational waveforms from compact binaries based on automated machine learning (AutoML) from around a hundred different possible regression models, without having to resort to tedious and manual case-by-case analyses and fine-tuning. The particular study of this article is within the context of the gravitational waves emitted by the collision of two spinless black holes in initial quasi-circular orbit. We find, for example, that approaches such as Gaussian process regression with radial bases as kernels do provide a sufficiently accurate solution, an approach which is generalizable to multiple dimensions with low computational evaluation cost. The results here presented suggest that AutoML might provide a framework for regression in the field of surrogates for gravitational waveforms. Our study is within the context of surrogates of numerical relativity  simulations based on Reduced Basis and the Empirical Interpolation Method, where we find that for the particular case analyzed AutoML can produce surrogates which are essentially indistinguishable from the NR simulations themselves. 
\end{abstract}

\maketitle

\section{Introduction and Overview} \label{sec:intro}

Construction of real-time or near real-time gravitational wave surrogates based on a small but carefully selected set of known waveforms has over the last years become a discipline on itself in gravitational wave (GW) science, for a review see \cite{tiglio2021reduced}. These surrogates are based on a {\it fiducial} or underlying model, which can be obtained from full numerical relativity (NR) solutions to the Einstein equations or physical approximants such as Effective One Body (EOB), Phenomenological (Phenom) or Post-Newtonian (PN) ones, as physical examples. In the reduced basis (RB)~\cite{prud'homme:70,Field:2011mf} approach a set \TSrb{} of fiducial waveforms is used to build a compact basis which approximates it with high accuracy. In some but not all cases the approximation error is smaller than that one of the fiducial waveforms themselves, it is in this sense that we refer to  the surrogate model as being essentially indistinguishable from the fiducial one. For NR the error of the fiducial model can refer to truncation error when solving set of elliptic-hyperbolic equations that constitute Einstein's theory, and in the case of approximants to the systematic error present due to physical approximations.  

\begin{figure}[h!]
    \centering
    \includegraphics[width=0.55\textwidth]{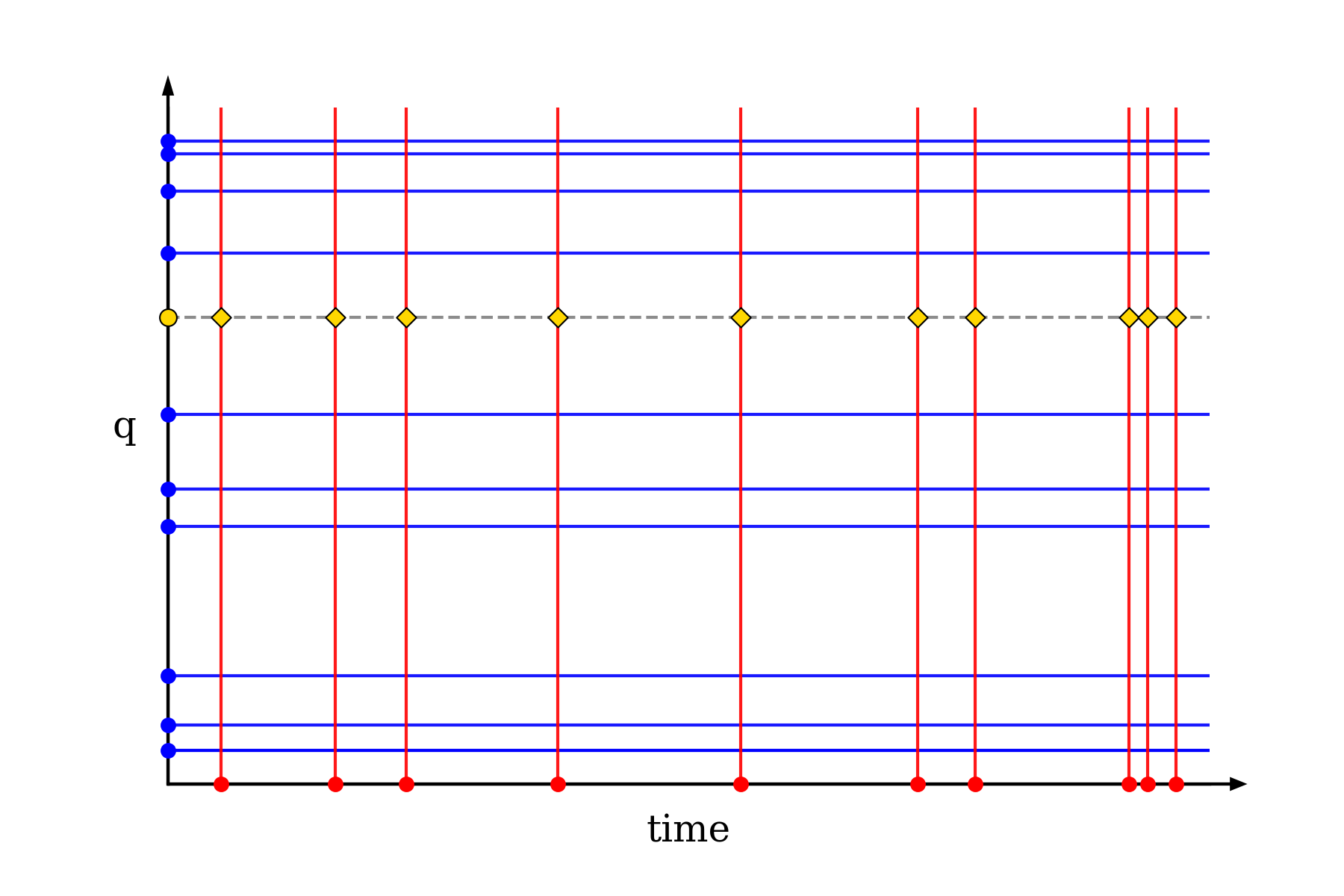}

    \caption{Schematic representation of the three steps surrogate pipeline. {\bf Horizontal blue} lines represent the reduced basis obtained by the greedy algorithm in step~1 ({\bf blue dots} are the greedy parameters of each basis element). {\bf Vertical red} lines represent the regressions obtained in step~3 ({\bf red dots} are time nodes obtained in step~2). The {\bf horizontal dotted} line is a predicted wave at parameter $q$ ({\bf yellow circle}) just obtained by evaluating regressions ({\bf yellow diamonds}) at $q$. }
    \label{fig:schematic}
\end{figure}

One approach for building surrogates, which as of this writing is the one commonly used when NR is the fiducial model, is through the following three steps  \cite{PhysRevX.4.031006}:

\begin{enumerate}
\item A sparse {\it reduced basis} with $n$ elements is chosen by a special selection of waveforms from \TSrb{}, of size $N$, through a greedy algorithm, until a sufficiently small user-specified tolerance error $\epsilon$ is met. Then, the number of instances is reduced from $N$ waveforms to $n$ with, in general, $n \ll N$.

\item A sparse subset of time (or frequency) nodes is chosen using the Empirical Interpolation Method (EIM), from $L$ time samples to $l \ll L$. The number  $l$ of EIM nodes is by construction the same as the number of greedy parameter ones: $l = n$. This therefore achieves a second compression, now in the time (frequency) domain, on top of the compression in parameter space from the RB-greedy first step, leading to an overall compression factor of $(L \times N)/n^2 \, $.

\item The previous two steps provide a {\em representation} of {\em known} waveforms. The next goal is to build a {\em predictive} surrogate model. That is, one which can produce new waveforms not present in the training set \TSrb{}. This is achieved performing regressions on waveform values at each EIM node. The newly predicted waveform is then obtained at all times using the EIM. For a summary of all these steps combined see Figure~\ref{fig:schematic}.

\end{enumerate}

With steps 1-2 one can achieve very low representation errors. This leaves the third step as the principal source of error in the pipeline outlined above (see Section~\ref{sec:feat-eng}), which motivates us to search for methods that contribute the less possible to the error of the surrogate models. Regression is usually performed using polynomials through interpolation or least squares. Even though polynomials approximations are a seemingly natural choice for which there is a vast amount of theory, they could suffer from several drawbacks in high dimensional problems, such as the curse of the dimensionality~\cite{Bellman:DynamicProgramming, 2017arXiv170306987A, CHKIFA2015400} and Runge's phenomenon~\cite{10.2307/2323093}, the latter becoming even more relevant when data are sparse and a tradeoff between stability and precision is needed~\cite{Cohen2013OnTS, cohen2017}.
Then, it is worth looking at non-algorithmic (probabilistic) approaches to regression. Here we are interested in approaches from Machine Learning (ML), for which there is a huge diversity.

AutoML~\cite{automl_book} is a research and development area of Artificial Intelligence (AI) that targets the end-to-end automation of the machine learning process. It allows the use of ML  techniques of any complexity in order to solve problems without full expert knowledge. AutoML intends to automate the learning workflow in all its stages: data preparation (e.g. feature extraction, transformation of non structured data to tabular form), feature engineering (missing and null data imputation, feature selection, feature generation), model selection and tuning (hyperparameter optimization), metric generation, data and result visualization.

In this work we analyze the possibility of using AutoML to drive the selection of sufficiently accurate regression approaches from $\sim 100$ options. We restrict our study to the case of regression at EIM nodes as needed when building surrogates.
 The physical scenario used as testbed is that one of the GWs emitted by the collision of two black holes initially without spin and in quasi-circular orbit, as predicted by numerical relativity. We used DataRobot~\cite{datarobot}, an AI platform which provides an AutoML pipeline for automation of all steps described above along the usual ones such as training, validation and testing. We address feature engineering before automation through dimensional reduction techniques. The idea is to build synthetic features which might further improve the learning process.

We find that out of $\sim 100$ set of available models, the most accurate surrogates are built based on three approaches: Gaussian Process Regression (GPR), Support Vector Machines (SVM) and Symbolic Regression through Genetic Programming (Eureqa).

This article is organized as follows. In Section~\ref{sec:setup} we discuss the physical and computational setups for our study, and in Section~\ref{sec:automl} the AutoML flow used. Section~\ref{sec:results} presents our results. Appendix~\ref{app:models-blueprints} lists the blueprints for AutoML used in this work and Appendix~\ref{app:rb-eim} describes the construction of our reduced basis and empirical interpolant, both of which are used to build features for AutoML. We close with some remarks putting this work in perspective in Section~\ref{sec:discussions}. 

\section{Setup} \label{sec:setup}

\subsection{Fiducial Models}

The physical system used as testbed in this article consists of waveforms emitted by the collision of two black holes initially without spin and in quasi-circular orbit, in the time range $[-2750,100]$M, where M is the total mass and the waveforms are aligned so that $t=0$ corresponds to the peak of the amplitudes of the GWs (approximately, the time of merger of the two black holes). This setup corresponds to $\sim 15$ orbits before merger. Due to the scale invariance of GR, the only relevant parameter is the mass ratio $q:=m_1/m_2$ -- with $m_i$ the mass of each black hole -- here chosen to be in the range $q \in [1,10]$. For definiteness we restrict our studies to the dominant $\ell = m = 2$ multipole angular mode.

As points of comparison we use two almost equivalent fiducial models for our physical scenario: 
\begin{enumerate}
\item High accuracy full NR simulations of the Einstein equations for $22$ different values of mass ratio $q$. The associated waveforms are summarized in Table I of Ref.~\cite{PhysRevLett.115.121102}, publicly available from the SXS Gravitational Waveform Database~\cite{sxs}. We generically denote a NR waveform as \hNR{}.  

\item A RB-EIM surrogate built in~\cite{PhysRevLett.115.121102} following the prescriptions described in the previous section, from those  $22$ NR simulations, using polynomial least squares for the regression step. The model agrees after a simulation-dependent time shift and physical rotation with the full NR results within numerical truncation errors with the advantage of being fast to evaluate for any value of $q\in[1, 10]$. We refer to the associated waveforms as polynomial least-squares or least-squares (for short) waveforms, and denote any of them as \hLS{}. We compute them using the publicly available Python package GWSurrogate~\cite{gwsurr}. 
\end{enumerate}

In this work we present an  automatic construction of a myriad of waveform surrogates using the three steps mentioned in the previous section but with a large number of ML regression approaches through AutoML. We generically refer to them, for short, as our ML surrogates, and the associated waveforms as \hML{}. There are several steps in this exploratory study in which the fiducial model needs to be a posteriori evaluated for any value of $q$ as many times as needed, in a very fast manner. One of these steps is in the testing process of the AutoML flow. 
Since we use a Cross Validation (CV) approach with a small and sparse training set, being able to do a thorough testing of the models with highest CV scores allows us to see if there are any correlations between those scores and test ones (we find that, indeed, there are). Next, for this study we need to evaluate the errors of ML models with respect to a fiducial reference. For these reasons we use the least-squares surrogate as fiducial model to build our ML ones. Not withstanding, we also measure the errors of ML models compared to numerical relativity waveforms.

\subsection{Feature engineering: Reduced Basis and the Empirical Interpolation Method}\label{sec:feat-eng}

Instead of training a complete regression pipeline with features coming from discrete waveform values $h(q_i; t_j)$ -- something that would pose a two dimensional problem for each waveform component, we build new features through dimensional reduction using the RB and EIM approaches. Here a waveform $h$ is thought of as a parametrized function of time $t$, the parameter $q$ is the mass ratio of the black holes.

In this context, the goal of the RB method is to build from a training set \TSrb{} of waveforms a basis able to span all training elements within a given tolerance (in this study, machine precision). This is done by iteratively selecting from the parameter space a hierarchical set of greedy points labeling those waveforms that will end up forming the reduced basis. The waveforms chosen to conform the basis are orthonormalized for conditioning purposes. 
Therefore, an RB expansion has the form
$$
h(q; t) \approx \sum_{i=1}^n c_i(q) e_i(t)\,,
$$
where $c_i$, given some inner product, is the projection coefficient of $h(q; t)$ onto the $i$-th basis element. The number of basis elements $n$ can be controlled by specifying a tolerance  $\epsilon$ for the representation error of the expansion. Lower tolerance means adding more basis elements to the actual basis to meet that precision.

On a second stage, the EIM allows us to build from the reduced basis an empirical interpolant of the form
\be
\cI[h](q; t) := \sum_{i=1}^n B_i(t) h (q, T_i)\, , 
\label{eq:interp}
\ee
which can also approximate any waveform in the training set with high accuracy.

The $T_i\,(i=1,\ldots,n)$ are the EIM nodes, a set of points especially selected from the time domain through a greedy procedure in the EIM. The functions $B_i(t)\,(i=1,\ldots,n)$ satisfy $B_i(T_j) = \delta_{ij}$ for all $i, j$ and are computed from the EIM nodes and the reduced basis. Later, both greedy and empirical samples (in parameter and time, respectively) play a prominent role in our regressions. A more extended review with applications to GW surrogates can be found in~\cite{tiglio2021reduced}.

Additionally, we decompose waveforms into phase and amplitude
$$
h(q; t) = A(q; t) e^{i\phi(q; t)}\,,
$$
each of which has less structure than the corresponding bare complex waveforms $h$, therefore facilitating the learning process. In this manner, we build a set of functions
\be
\Big\{
\phi_i(q) := \phi(q; T_i),\, A_i(q) := A(q; T_i)
\Big\}_{i=1}^n
\label{eq:synthetic}
\ee
composed by $2\times n$ phase and amplitude functions of $q$ evaluated at the EIM nodes that will serve to select the relevant features for regressions.

In the regression step we build surrogates for each $\phi_i(q)$ and $A_i(q)$, where $i$ labels the corresponding EIM node $T_i$. We denote them as
\be
\phi^{\text{ML}}_i(q),\,
A^{\text{ML}}_i(q)\quad (i=1,\ldots, n)\,.
\label{eq:regs}
\ee
In the next section we discuss the selection of $q$ points to conform the training set for this step.

Inserting regressions for phases and amplitudes into the interpolant formula (\ref{eq:interp}) leads us to the assembly of the final surrogate:
\be
h_{\text{ML}}(q; t) := \sum_{i=1}^n B_i(t) A^{\text{ML}}_i(q) e^{i\phi^{\text{ML}}_i(q)} \,.
\label{eq:surr}
\ee
In the previous section we remarked that the regression step is the most important source of error in building predictive models once the RB-EIM representation is accurate enough. We can see this from the next inequality for discretized time~\cite{PhysRevX.4.031006}:
$$
\Delta t \sum_{i=1}^{L}\vert h(q ; t_{i})-h_{\text{ML}}(q ; t_{i})  \vert^{2} \leq \Lambda_{n} \sigma_{n}+\Lambda_{n} \Delta t \ \sum_{i=1}^{n} \vert h(q ; T_{i})-h_{\text{ML}}(q ; T_{i}) \vert^{2}\,.
$$
The first term in the r.h.s. is the error associated to the RB-EIM approximation, which we ensure is small enough (at the level of machine precision) for all practical purposes. The second term corresponds to regression and is the leading source of error.

\subsection{Training method for regressions}

The selection of a suitable training set \TSreg{} for phase and amplitude regressions comes from a compromise with data availability. Indeed, when the surrogate is to be built from NR, the number of available simulations is very limited due to the computational cost of numerical evolving the full Einstein equations. 

As an example, the first NR GW surrogate (non-spinning binary black holes, which is the test case discussed in this article) used an EOB approximation to select the greedy points in parameter space, and afterwards full NR simulations were ran for those values. That is, there was no training set of NR waveforms to build the basis but, instead, one based on EOB. Similarly, for the full 7-dimensional case, as of this writing $\sim 1,000$ NR waveforms are used as training set to build a reduced basis with a compression factor smaller than $\sim 2$.  So, as a worse case scenario one might consider that only the reduced basis waveforms are available. In other scenarios, where it is not so expensive to produce those waveforms, such as in EOB, Phenom or PN approximations, one might have a dense training set to start with.

Following the above worse case rationale, we choose to conform the \TSreg{} with phase and amplitude values at greedy points and EIM nodes obtained from the application of the RB and EIM algorithms to the \TSrb{}.
We summarize the results from this step in Appendix~\ref{app:rb-eim}.

The application of these algorithms using a tolerance equal to machine precision ($\epsilon = 10^{-16}$) gives us a total of $26$ greedy parameters and the same number of EIM nodes. In consequence, the training set for regressions takes the form 
\be
\text{TS}_\text{reg} = \{\phi_i(Q_j),\,A_i(Q_j) \}\,, \qquad (i,j=1, \ldots, 26) \, ,  \label{eq:ts-reg}
\ee
where $Q_j$ denote the greedy parameter points and $i$ labels the EIM nodes.
In other words, to obtain a single surrogate model (\ref{eq:surr}) the $26 \times 2 = 52 $ phase and amplitude regressions (\ref{eq:regs}) as functions of $q$ need to be calculated at each $T_i$.

There are two possible data partitioning approaches for training an AutoML pipeline and comparing regression methods: i) train, validation (TV)  and ii) $k$-fold cross validation (CV). A  test set is also needed for either approach, which is never seen in training stages and is used to give a final estimation of models' accuracy for unseen samples. In our case it has data corresponding to 500 waveforms.

With TV, the entire dataset is split into three parts: training, validation and test. ML models are trained in the first set and validated in the second one. In general, this strategy is well suited for large datasets for which there is less bias induced by a train-validation split. On the other hand, leaving apart the test set, $k$-fold CV splits the data into smaller subsets (folds) and models are trained $k$ times in $(k-1)$ folds with the remaining one left for validation. At the end of the process, all data points are used for training, thereby reducing bias. 

Since our datasets are small, we choose a special case of $k$-fold CV as our training strategy: leave-one-out cross-validation (LOOCV). In this case, the number of partitions equals the number of training samples. Since each regression dataset consists of $26$ points, we set $k=26$.

\subsection{Surrogates assessment}

In Section~\ref{sec:results} we compare our ML-regression surrogates $h_{\text{ML}}$ with fiducial waveforms $h_f$.
Here, $h_f$ is a numerical relativity $h_{\text{NR}}$ or a least-squares waveform \hLS{}. 
For those comparisons we use the relative $L_2$ error, 
\be
E = E(q) := \frac{\left \lVert  h_{\text{ML}}(q,\cdot) - h_f(q,\cdot) \right \rVert^{2}}{ \left \lVert h_{f}(q,\cdot) \right \rVert^{2}} \ , \label{eq:rel-error}
\ee
where
$$
\left \lVert h(q,\cdot) \right \rVert^{2}  =   \int dt\  \vert h(q,t)\vert ^{2}\, . \label{eq:norm}
$$
For computation of integrals we use Riemann quadratures with $\Delta t = 0.1$M.

\section{AutoML flow} \label{sec:automl}

DataRobot (DR) is an AutoML platform that applies ``best practices'' in the ML process leveraging a high level of automation to build efficient ML models. For modeling, DR uses special pipelines called blueprints. Each blueprint maps inputs to predictions by concatenating a ML model with a preprocessing step. For the datasets used in this study DR uses $\sim 60$ different ML models which come from open source and proprietary algorithms such as Eureqa~\cite{eureqa}, H20~\cite{h20}, Keras~\cite{keras}, LightGBM~\cite{lgbm}, Scikit-Learn~\cite{scikit-learn}, TensorFlow~\cite{ts}, Vowpal Wabbit~\cite{vowpalwabbit}, and XGBoost~\cite{Chen2016XGBoostAS}. These models are combined with several preprocessing methods to finally conform the  $87$ blueprints described in Appendix~\ref{app:models-blueprints}. This, and the possibility of creating ensembles of different models, give us a wide spectrum for AutoML exploration. We point out that many of the underlying ML algorithms used by DR are open-source and available, for example, in Scikit-Learn.

\blank
\noindent{\bf Data Partitioning}.-- %
The first stage is to split data into train and test partitions. The test partition is used to compute a final estimation of models' generalization.
The second stage consists in splitting the train partition into $k$ folds for subsequent application of $k$-fold CV.
Recall that in this work $k=26$ (LOOCV).

Below we illustrate how the partitions are arranged in a generic $k$-fold CV approach.

\begin{figure}[h!]
\centering
 \includegraphics[width=0.6\textwidth]{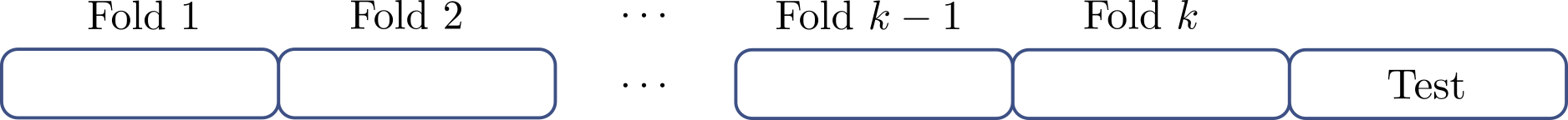}
\end{figure}

\blank
\noindent{\bf Training and hyperparameter tuning}.-- %
In an AutoML flow we can trace at least three optimization levels. The first optimization relies on selecting the best set of hyperparameters for each model, that is, parameters that control the behavior of the ML algorithms. In a Deep Neural Net, for example, the hyperparameters correspond to the number of layers (depth), the number of units per layer, the learning rate and penalizing constants, just to name the most relevant. This process is called hyperparameter tuning. Completed the first level, the next step is to train each model with the optimized hyperparameters. The final step is to rank all models following some metric, in this study, Root Mean Square Error (RMSE). 

For this work, hyperparameter tuning is performed through grid searches using $k$-fold CV over training datasets. Below we illustrate the process for a grid of $n$ samples. A dataset is divided into $k$ partitions and for each hyperparameter tuple corresponding to some ML algorithm, $k$ models are trained, each of one using $(k-1)$ different partitions. As usual, predictions are computed for each model in the remaining partition.

\begin{figure}[h!]
\centering
 \includegraphics[width=0.7\textwidth]{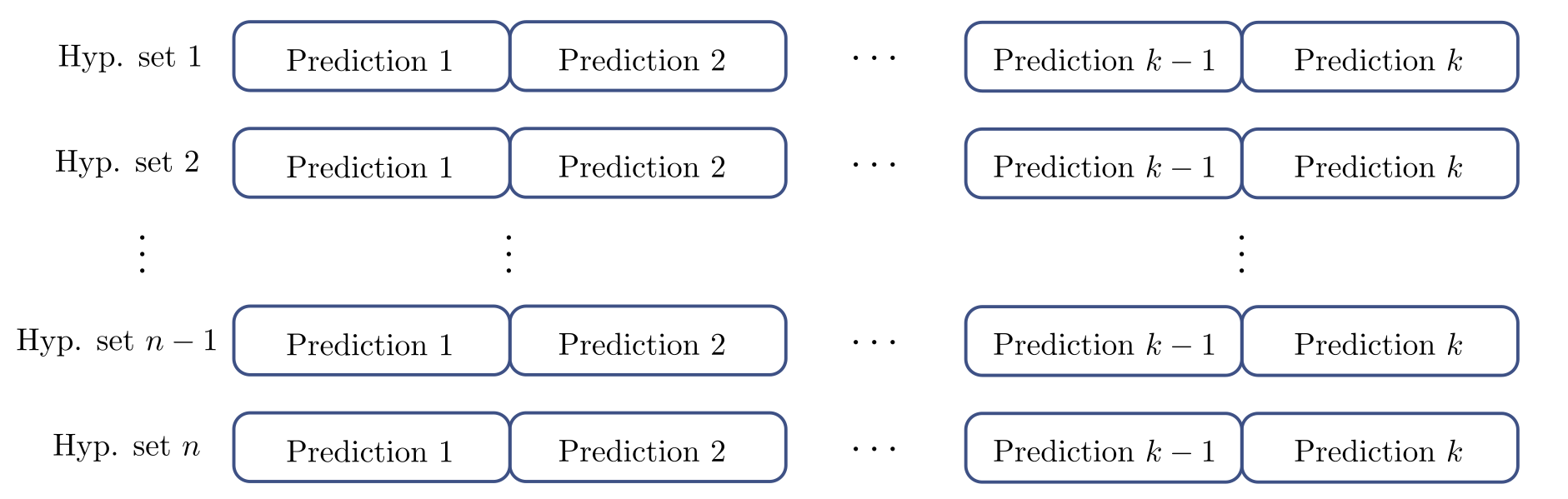}
\end{figure}

The remaining $k$ folds cycle through the train-validation process ($k$ times, $k$ validation scores) and then a final CV score is computed as an average.
At the end, models are prepared for final training with the entire $k$-folds' data by selecting the best hyperparameters for it.

\blank
\noindent{\bf Model scoring}.-- %
All models are ranked according to their CV scores. This gives us a list of the most accurate ones with respect to the train partition. Finalized the training stage, the test partition is used for models assessment on unseen data.

\section{Results} \label{sec:results}

We now show the results for regressions as described in the previous two sections. We perform fits along waveforms phase and amplitude values at each EIM node by {\em only} using a sample in the parameter space corresponding to the $26$ RB greedy values of $q$ at that node for training.
Then we evaluate and rank their accuracy according to their LOOCV and test RMSE scores to choose the ones that perform best along training and test datasets, respectively, for each phase and amplitude groups.

From the AutoML perspective and this work each blueprint consists of an ML algorithm and a set of preprocessing steps. A model, in turn, is the output of training a blueprint. To have a short terminology for each blueprint we use BP 1, BP 2, ..., BP 87, in decreasing order according to the accuracy of their corresponding waveform surrogates, we refer to Table~\ref{tab:errors}. For accuracy we use the maximum RMSE over $500$ test values of $q$.
In Appendix~\ref{app:models-blueprints} we list all the blueprints used. 

We use the DR Python API~\cite{DR-API} to perform our regressions and the Python language for surrogates construction and validation.

\subsection{Regressions} 

We bench the performance of phase and amplitude regressions over the \TSreg{} by computing the mean LOOCV for each blueprint, and for each phase and amplitude groups of regressions.
For instance, in the case of phase regressions, we have $26$ phase models (one per EIM node) per blueprint, each of which has its own LOOCV score.
This gives us $26$ LOOCV scores that we average to quantify the overall performance of that blueprint across those phase regressions.
This is done for each blueprint inside the phase group of regressions. We also replicate this process for amplitude regressions.

\begin{figure}[h!]
    \centering
\includegraphics[width=0.45\textwidth]{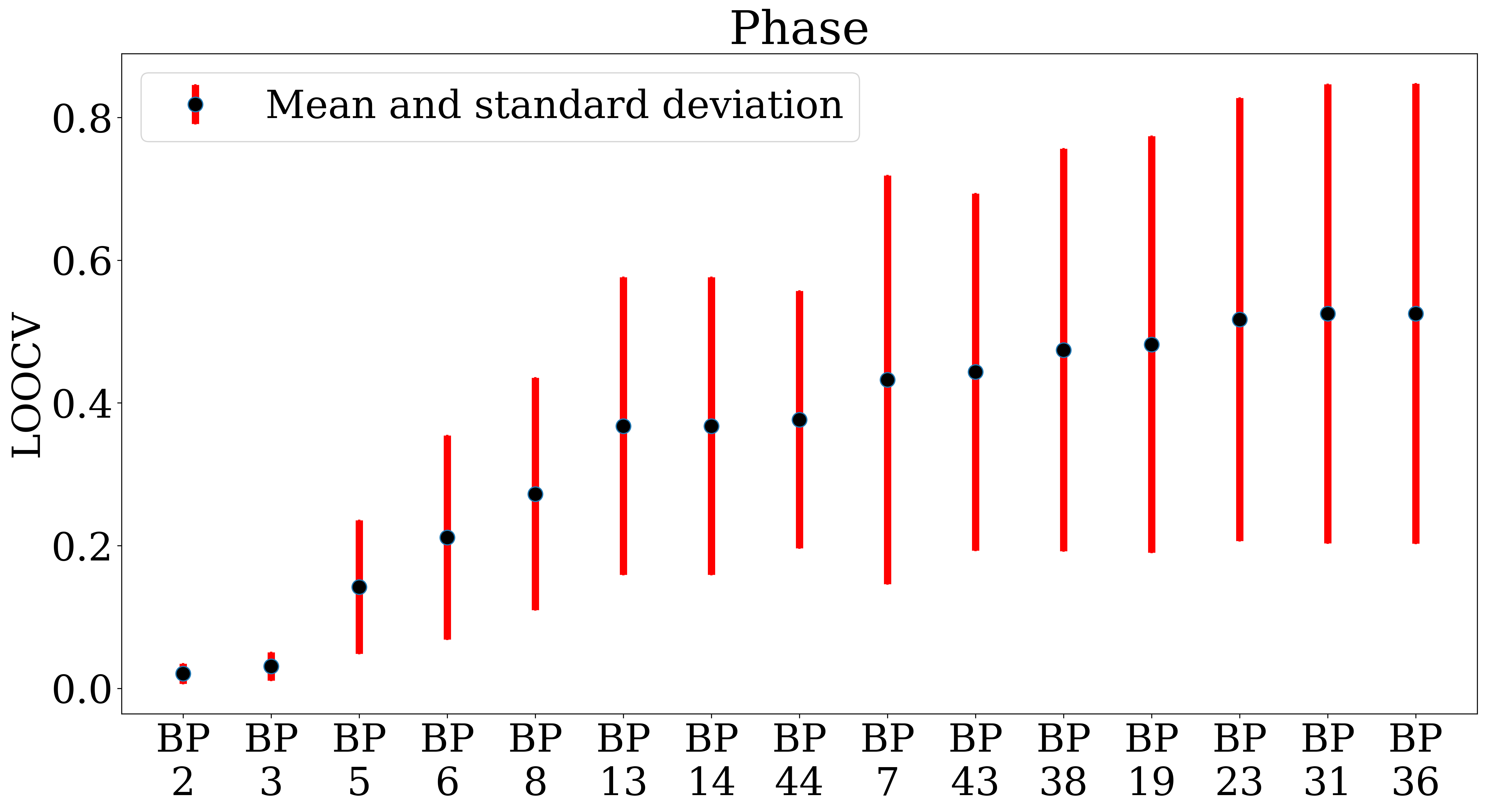}
\includegraphics[width=0.465\textwidth]{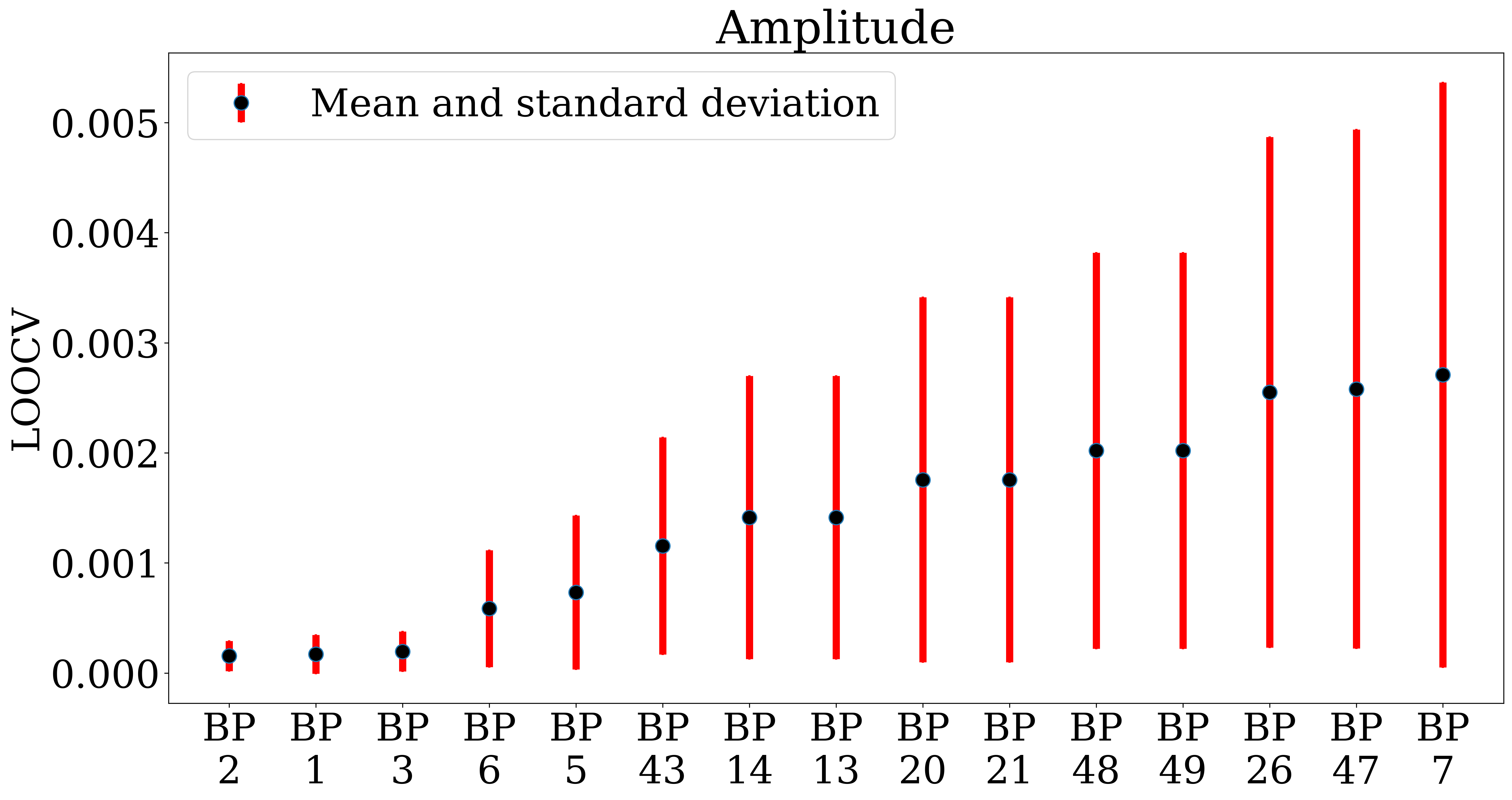}
\caption{First $15$ ranked models for phase and amplitude datasets in LOOCV, sorted by mean LOOCV score. Mean values (black dots) are displayed along standard deviations (red lines) of LOOCV scores per BP. Note: Eureqa models are not included in this ranking since the AutoML platform used does not give LOOCV scores for that particular model for technical implementation reasons.}
    \label{fig:loocv_max_validation}
\end{figure}

The results are shown in Figure~\ref{fig:loocv_max_validation}, from which one can see that there is a predominance of Gaussian Process Regression (GPR) models, ranked in first places. In particular, for phase and amplitude there is the same model (BP 2) ranked in first place, it is a GPR model with Mat\'ern kernel (see Appendix~\ref{sec:regression}).

\begin{figure}[h!]
    \centering
    \includegraphics[width=0.45\textwidth]{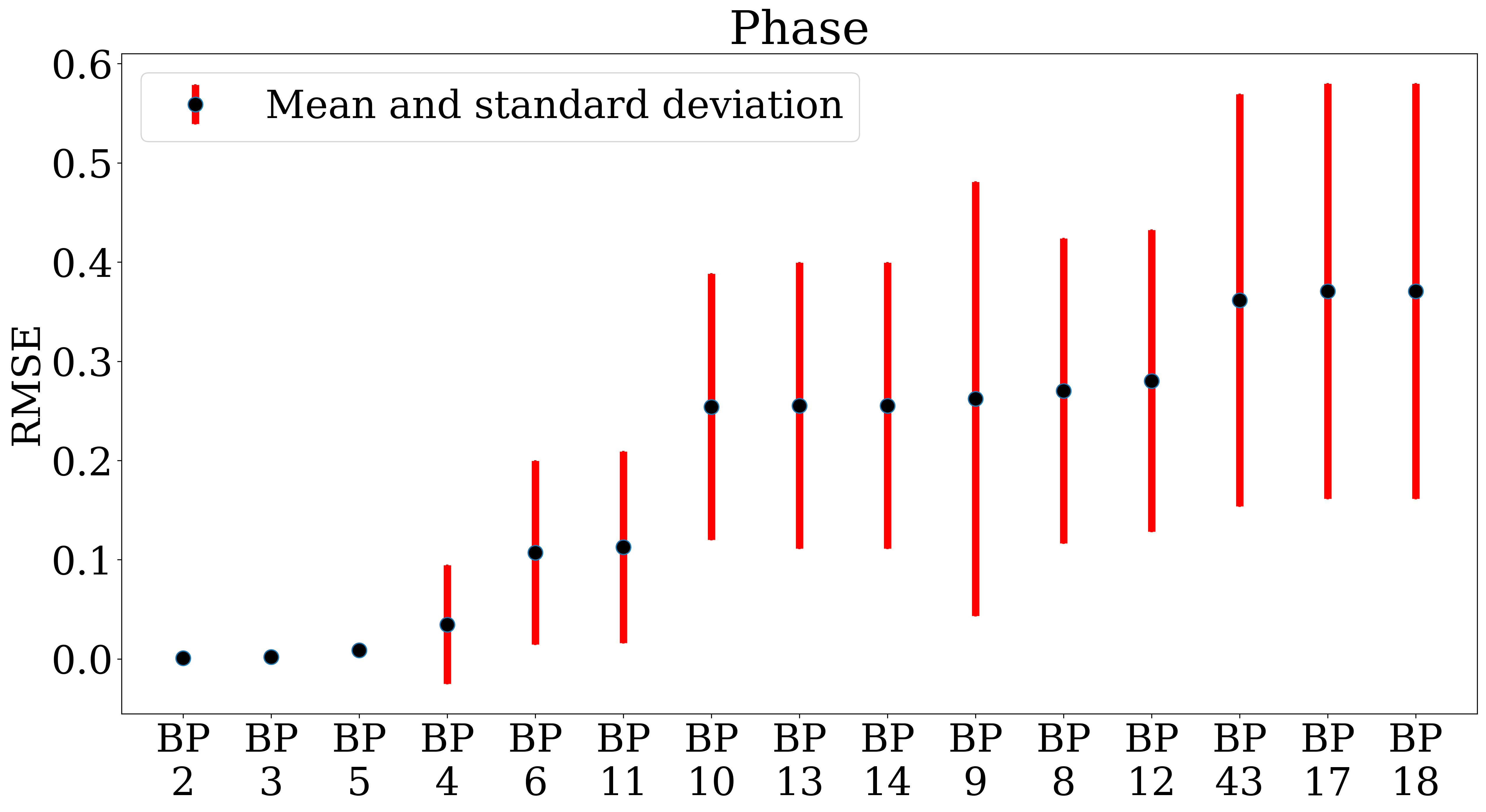}
    \includegraphics[width=0.47\textwidth]{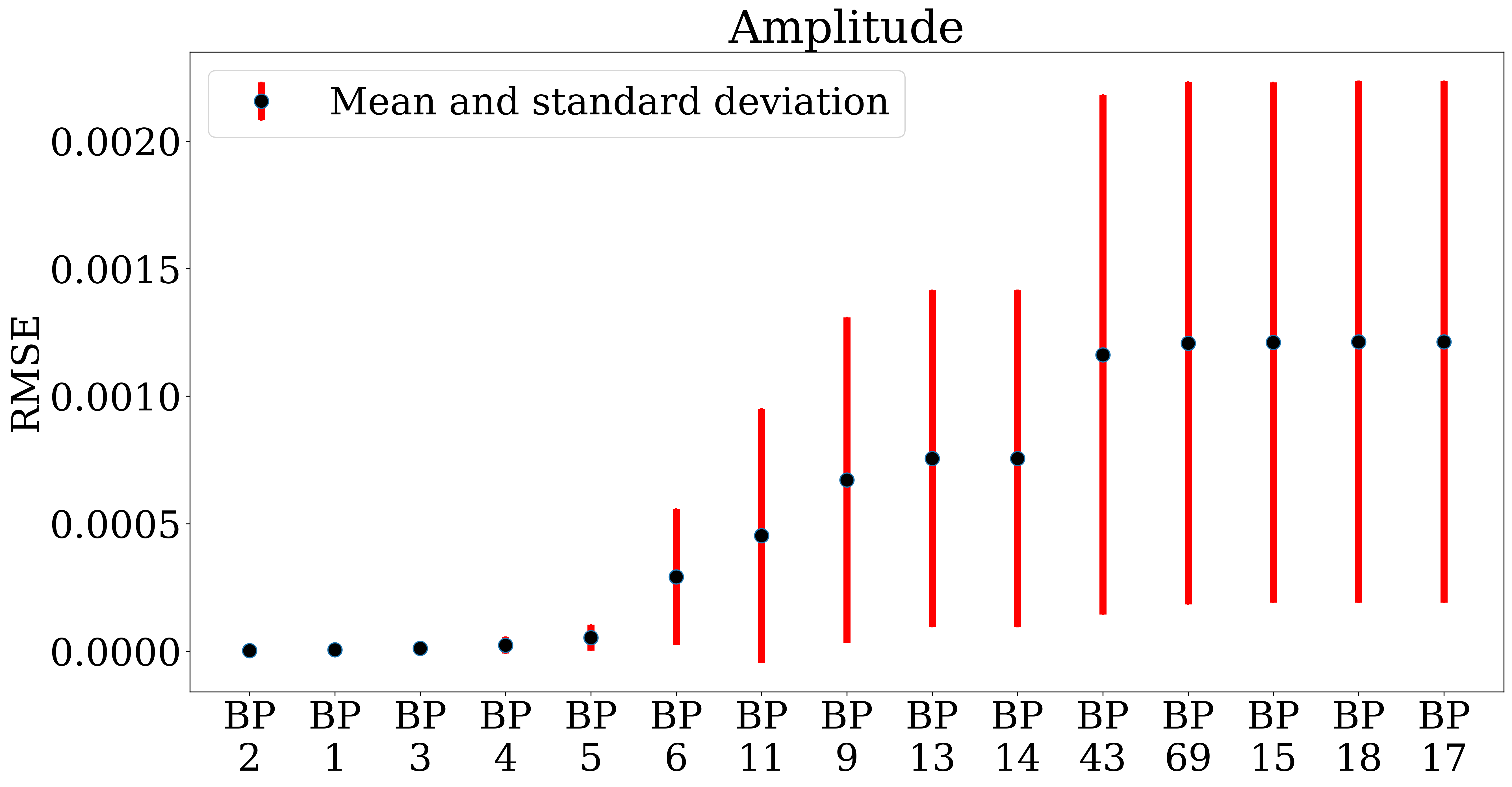}
    
    \caption{First $15$ ranked models for phase and amplitude datasets when testing, sorted by the mean values of RMSE with respect to $500$ test values of $q$.} 
    \label{fig:loocv_max_holdout}
\end{figure}

Figure~\ref{fig:loocv_max_holdout}, in turn, shows the equivalent results but now using the $500 \times 2$ test datasets to compute the RMSE scores. An observation is that blueprints 1-6 appear among the highest ranked $15$ ones both with respect to CV and test scores (with the exception of BP 4, Eureqa, for the reason explained in the figure).  This accordance cannot be related in any way to the ML methodology employed, since the CV score is calculated with the training dataset and the test score is independent of the regression model creation process. Furthermore, as we will see below, the models associated with these blueprints are also the ones with highest scores with respect to a completely different error metric, given by Equation~\ref{eq:rel-error}, for the whole surrogate waveforms.

\subsection{Waveform ML surrogates}  \label{sec:ml-surrogates}

\begin{table}[h!] \begin{tabular}{|c|c|c|c|c|} \hline

Max error rank & Max & Median & Mean \\ \hline 
BP 1 & 2.84e-07 & 6.71e-08 & 9.44e-08  \\ \hline

BP 2 & 2.86e-05 & 4.92e-11 & 3.19e-07  \\ \hline

BP 3 & 3.64e-04 & 2.90e-11 & 3.90e-06  \\ \hline

BP 4 & 4.68e-04 & 3.15e-05 & 3.35e-05  \\ \hline

BP 5 & 1.50e-03 & 2.17e-05 & 7.58e-05  \\ \hline

BP 6 & 1.08e-01 & 4.13e-03 & 8.33e-03  \\ \hline

BP 7 & 4.39e-01 & 1.73e-01 & 1.79e-01  \\ \hline

BP 8 & 7.72e-01 & 7.05e-02 & 1.02e-01  \\ \hline

BP 9 & 9.16e-01 & 3.66e-01 & 3.81e-01  \\ \hline

BP 10 & 9.89e-01 & 2.10e-01 & 2.28e-01  \\ \hline

BP 11 & 1.01e+00 & 3.31e-02 & 5.17e-02  \\ \hline

BP 12 & 1.08e+00 & 2.14e-01 & 2.47e-01  \\ \hline

BP 13 & 1.10e+00 & 3.03e-04 & 8.76e-02  \\ \hline

BP 14 & 1.10e+00 & 3.03e-04 & 8.76e-02  \\ \hline

BP 15 & 1.10e+00 & 8.26e-02 & 1.87e-01  \\ \hline 
\end{tabular} 
\caption{Results for different ML surrogate models, sorted by maximum (over the $500$ test values of $q$) relative error (\ref{eq:rel-error}).} 
\label{tab:errors}
\end{table}

We have so far analyzed the ranking of the $15$ most accurate blueprints for regression of the $26\times 2$ phase and amplitude datasets according to LOOCV and test RMSE scores. We now turn to how well different blueprints perform on the resulting complex surrogate waveforms themselves, since those are the ultimate functions of interest. 

For this purpose we constructed $87$ ML surrogates as described in Sections~\ref{sec:intro} and ~\ref{sec:feat-eng}. For each surrogate we then computed the $500$ relative waveform errors defined by Eq.~(\ref{eq:rel-error}) using the test values of $q$ and least-squares surrogate \hLS{} as fiducial model. The results are summarized in Table~\ref{tab:errors} for the $15$ models with highest accuracy. Out of them, the first six correspond to blueprints of three families: Gaussian Process Regression (GPR), Eureqa (symbolic regression) and Support Vector Machines (SVM) and are within the $15$ models with highest accuracy in terms of RMSE for both LOOCV and test scores, for phase and amplitude. In Appendix~\ref{sec:regression} we briefly describe each of these three algorithms. 

\begin{figure}[h!]
    \centering
\includegraphics[width=0.8\textwidth]{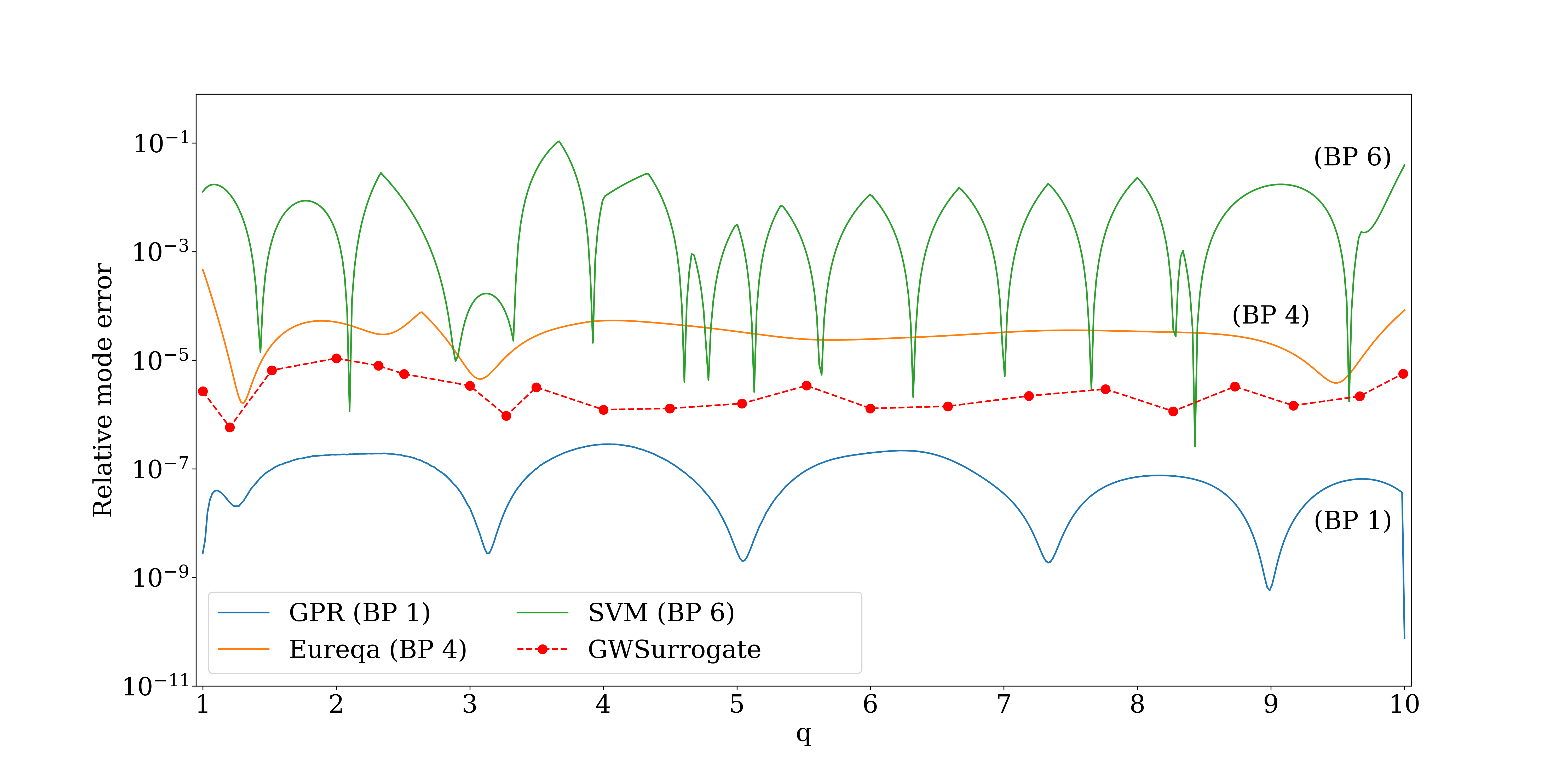}
    \caption{Relative waveform error (\ref{eq:rel-error}) for the three ML surrogate family models with highest accuracy. Blueprint 1 has errors which are below the errors of \hLS{} with respect to NR, which in turn are comparable to the truncation errors of NR.  
    } 
    \label{fig:err_automl}
\end{figure}

We next discuss in more detail the results for these three families. In Figure~\ref{fig:err_automl} we show the models with highest accuracy for GPR, Eureqa and SVM. The solid curves show the errors (\ref{eq:rel-error}) as a function of $q$, using the $500$ test values computed from the least-squares surrogate model \hLS{}. The dotted dashed red curve in turn shows the error of \hLS{}, here defined as (\ref{eq:rel-error}) where $h_f = h_{\text{NR}}$. That is, those $22$ red points show the error of least-squares surrogate waveforms with respect to the numerical relativity ones used to build it. We computed them using the package GWSurrogate and the public data for those $22$ NR simulations.
They have been shown in Ref.~~\cite{PhysRevLett.115.121102} to be comparable to the NR numerical truncation errors themselves.
We perform time and phase shifts on the NR waveforms using the Python Package GWTools~\cite{gwtools} so as to align them with \hLS{} waveforms.
Along this line here we notice that, for example, the errors of BP 1 with respect to \hLS{} model are in turn smaller than those of \hLS{} with respect to NR.

\begin{figure}[h!]
    \centering
\includegraphics[width=0.49\textwidth]{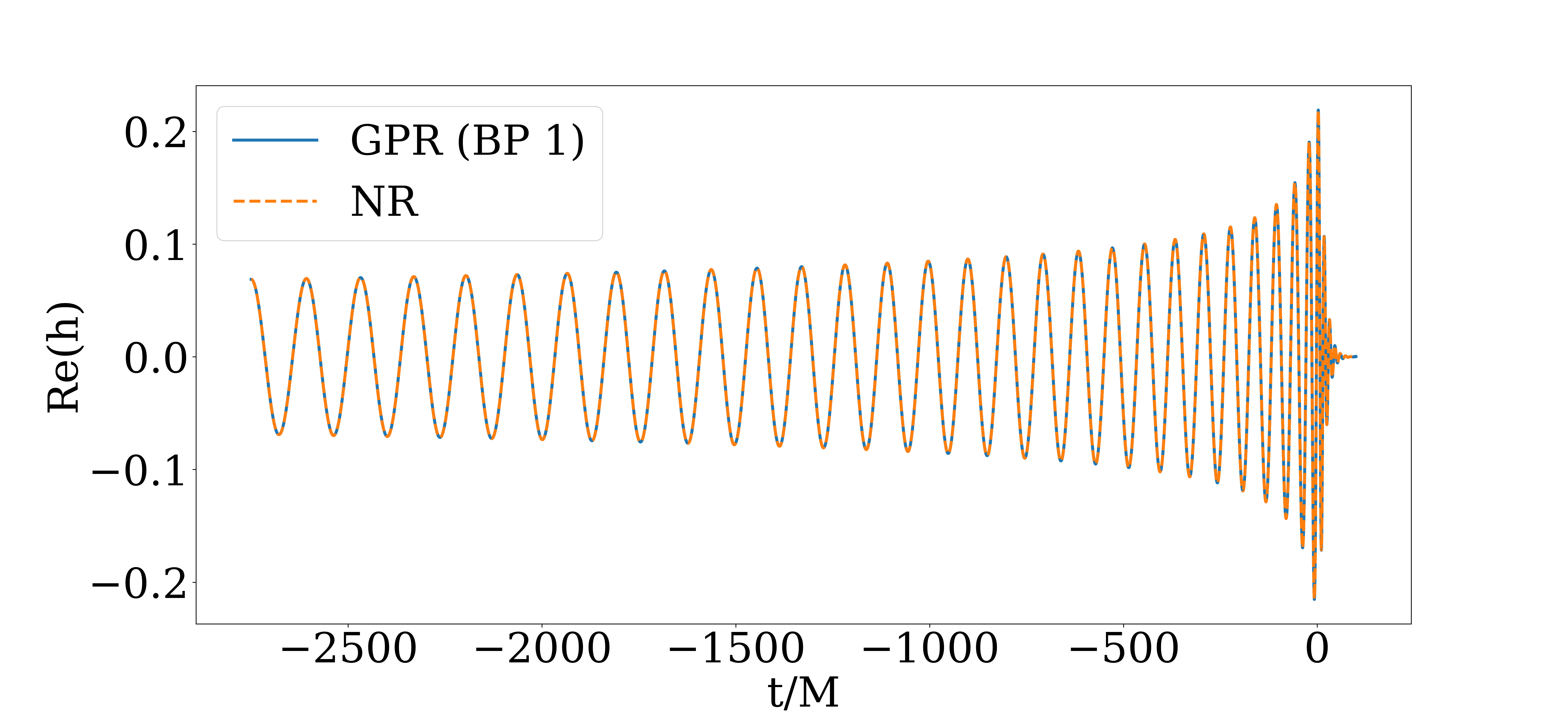}
\includegraphics[width=0.49\textwidth]{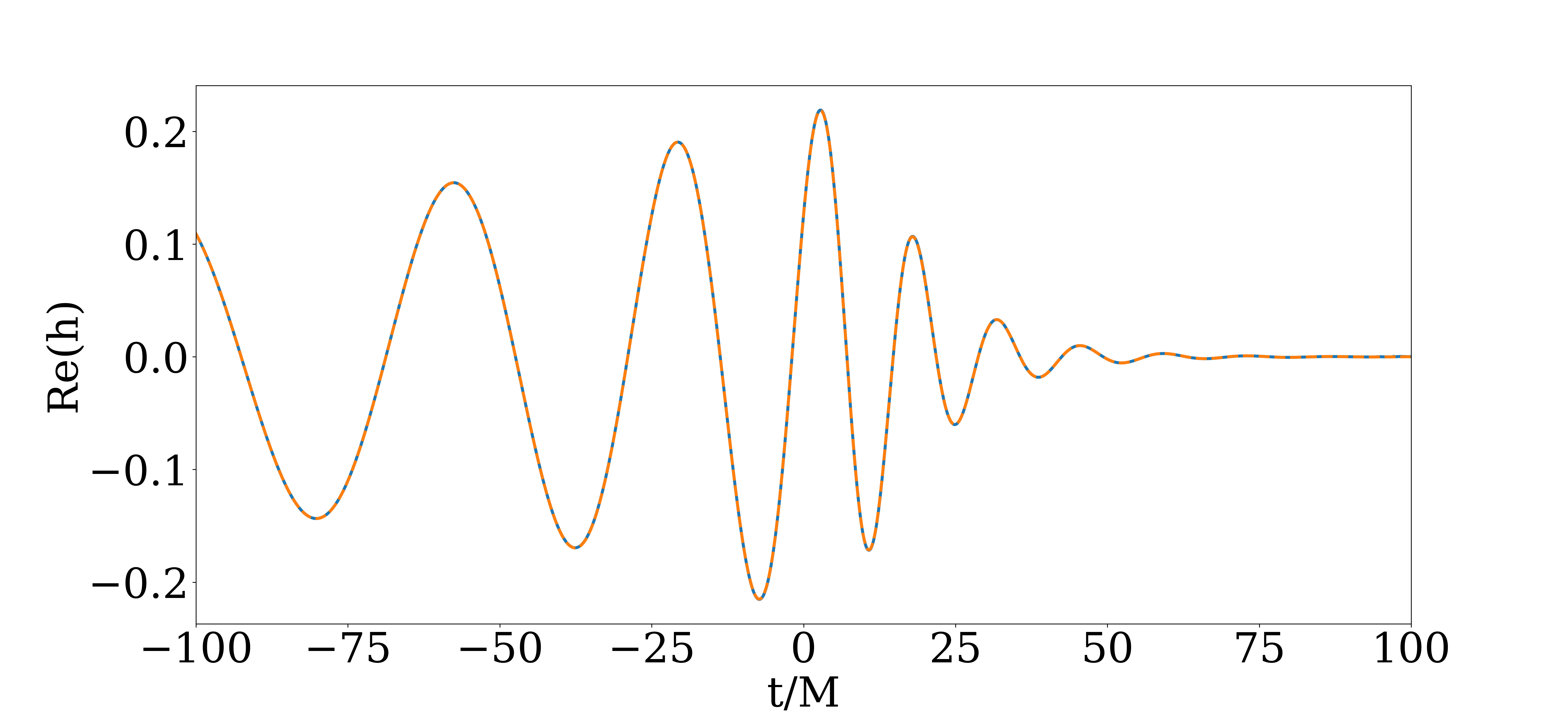}
    \caption{Prediction from the surrogate model with GPR and Radial Basis Function compared to the numerical relativity gravitational waveform for $q=4.499$.}
\label{fig:GPR_RB_VS_NR} 
\end{figure}

Figure~\ref{fig:GPR_RB_VS_NR} shows a comparison between the model resulting from BP 1 and NR waveform (plus polarization), for $q=4.499$. This is one of the $22$ NR values, close to the center of the interval used for $q$:  $[1,10]$.

\blank

\section{Discussions} \label{sec:discussions}

In this work we presented a case study in waveform modeling showing that, within the Reduced Basis-Empirical Interpolation (RB-EIM) framework, automated machine learning (AutoML) regression can help obtain waveforms which are very fast to evaluate and essentially indistinguishable from those generated by NR simulations.

This study is driven by two main motivations related to data availability and to harnessing the power of machine learning methods in the most autonomous way. 
The first one concerns the general issue of efficiently generating NR waveforms, as they are the endpoint of complex computational pipelines involving the numerical resolution of the Einstein equations.
The second motivation involves the reduction of bias in the learning process. Given some data distribution, the act of selecting any supervised ML algorithm to be trained on that data has an unavoidable inductive bias associated with the same selection process. Furthermore, the No Free Lunch Theorem~\cite{Wolpert1996TheLO} basically shows that there is no universal regressor capable of approximating equally well all possible data distributions. Then, the selection of an ML algorithm introduces severe constraints in learning since the possibility of any better regression approach is automatically discarded.
This fact motivates the use of AutoML to explore the most accurate possibility between available algorithms at a time of suppressing or, at least, diminishing human bias in the learning process. With AutoML we let available algorithms compete with each other for better performance, therefore avoiding potentially suboptimal decisions driven by human-biased criteria. 
Out of the $87$ regression blueprints analyzed by AutoML, we found that those based on Gaussian Process Regression, Symbolic Regression through Genetic Programming, and Support Vector Machines give the most accurate models (see Figs.~\ref{fig:loocv_max_validation}, \ref{fig:loocv_max_holdout} and \ref{fig:err_automl}).

We point out the high precision of these models regardless of the small size of datasets used for training, which were synthesized using the RB-EIM framework. In  order to test if this form of feature selection helps the ML algorithms reach their performance, we repeated all our experiments using random $q$ parameters instead of greedy points.
We found that the resulting surrogates can get to perform ~10-100 times worse than the greedy-based ones. 

Deep Learning-based models are not among the best performing ones in this study. Indeed, the Neural Nets blueprints used in this work rank last in the list of models arranged by performance. This can be related to the fact that deep architectures need in general large datasets to perform well. Since our datasets are small, their size could be insufficient for Neural Nets to reach good performance.

GPR-based models show a promising avenue for modeling waveform sparse data based on RB-EIM. The training cost of these models naively grows as $\bigo{n^3}$, with $n$ the size of the training set, although this does not pose a problem in this context given the small size of the training set.
This, and the linear cost por prediction once GPR models are trained makes these algorithms appealing for further research. Indeed, for batch sizes $\sim 10^{3}-10^{4}$, the cost per point evaluation of a GPR regression trained in this work stabilizes about fractions of milliseconds on a desktop computer.

To our knowledge this is the first regression study in waveform modeling using the RB-EIM framework and AutoML techniques.
A related work is~\cite{setyawati2019regression},
in which a few regression techniques for GWs are explored in the case of spin-aligned non-precessing and fully precessing compact binaries, although neither RB-EIM nor AutoML are used in the process, and the results have been claimed as being general knowledge without a comprehensive study~\cite{Cuoco_2020}. Our approach proposes a way of evaluating paradigms without human biases.

\section{Acknowledgments}
We thank DataRobot for a free academic license, William Disch, Scott E. Field and Ganesh Siddamal for helpful discussions.

\appendix
\section{Blueprints used} \label{app:models-blueprints}

List of the $87$ blueprints used in this study, ranked from smallest to largest error 
$
\max_{q \in [1,10]} E(q)
$  
with $E(q)$ defined by Equation~(\ref{eq:rel-error}).

\begin{itemize}
\item BP 1: Missing Values Imputed - Gaussian Process Regressor with Radial Basis Function Kernel
\item BP 2: Missing Values Imputed - Gaussian Process Regressor with Mat\'ern Kernel (nu=2.5)
\item BP 3: Missing Values Imputed - Gaussian Process Regressor with Mat\'ern Kernel (nu=1.5)
\item BP 4: Missing Values Imputed - Eureqa Regressor (Default Search: 3000 Generations)
\item BP 5: Missing Values Imputed - Gaussian Process Regressor with Mat\'ern Kernel (nu=0.5)
\item BP 6: Regularized Linear Model Preprocessing v20 - Nystroem Kernel SVM Regressor
\item BP 7: Numeric Data Cleansing - Standardize - Elastic-Net Regressor (mixing alpha=0.5 / Least-Squares Loss) with K-Means Distance Features
\item BP 8: Regularized Linear Model Preprocessing v22 with Unsupervised Features - Elastic-Net Regressor (mixing alpha=0.5 / Least-Squares Loss) with Unsupervised Learning Features
\item BP 9: Missing Values Imputed - Eureqa Regressor (Instant Search: 40 Generations)
\item BP 10: Tree-based Algorithm Preprocessing v1 - eXtreme Gradient Boosted Trees Regressor with Early Stopping
\item BP 11: Missing Values Imputed - Eureqa Regressor (Quick Search: 250 Generations)
\item BP 12: Tree-based Algorithm Preprocessing v20 - eXtreme Gradient Boosted Trees Regressor with Early Stopping
\item BP 13: Missing Values Imputed - Smooth Ridit Transform - Partial Principal Components Analysis - Gaussian Process Regressor with Mat\'ern Kernel (nu=0.5)
\item BP 14: Missing Values Imputed - Smooth Ridit Transform - Gaussian Process Regressor with Mat\'ern Kernel (nu=0.5)
\item BP 15: Tree-based Algorithm Preprocessing v1 - Adaboost Regressor
\item BP 16: Tree-based Algorithm Preprocessing v1 - Gradient Boosted Trees Regressor (Least-Squares Loss)
\item BP 17: Tree-based Algorithm Preprocessing v15 - Gradient Boosted Greedy Trees Regressor (Least-Squares Loss)
\item BP 18: Missing Values Imputed - Gradient Boosted Trees Regressor (Least-Squares Loss)
\item BP 19: Regularized Linear Model Preprocessing v19 - Ridge Regressor
\item BP 20: Regularized Linear Model Preprocessing v14 - Ridge Regressor
\item BP 21: Regularized Linear Model Preprocessing v14 - Ridge Regressor
\item BP 22: Numeric Data Cleansing - Standardize - Vowpal Wabbit Low Rank Quadratic Regressor
\item BP 23: Numeric Data Cleansing - Standardize - Ridge Regressor - Light Gradient Boosting on ElasticNet Predictions 
\item BP 24: Numeric Data Cleansing - Standardize - Vowpal Wabbit Stagewise Polynomial Regressor
\item BP 25: Numeric Data Cleansing - Standardize - Vowpal Wabbit Regressor
\item BP 26: Missing Values Imputed - Standardize - Auto-tuned K-Nearest Neighbors Regressor (Euclidean Distance)
\item BP 27: Tree-based Algorithm Preprocessing v1 - eXtreme Gradient Boosted Trees Regressor
\item BP 28: Tree-based Algorithm Preprocessing v20 - eXtreme Gradient Boosted Trees Regressor
\item BP 29: Tree-based Algorithm Preprocessing v22 with Unsupervised Learning Features - eXtreme Gradient Boosted Trees Regressor with Unsupervised Learning Features
\item BP 30: Numeric Data Cleansing - Standardize - Auto-tuned Stochastic Gradient Descent Regression
\item BP 31: Numeric Data Cleansing - Standardize - Lasso Regressor
\item BP 32: Numeric Data Cleansing - Standardize - Elastic-Net Regressor (mixing alpha=0.5 / Least-Squares Loss)
\item BP 33: Missing Values Imputed - Standardize - Elastic-Net Regressor (mixing alpha=0.5 / Least-Squares Loss)
\item BP 34: Missing Values Imputed - Standardize - Ridge Regressor
\item BP 35: Missing Values Imputed - Standardize - Linear Regression
\item BP 36: Numeric Data Cleansing - Standardize - Ridge Regression
\item BP 37: Tree-based Algorithm Preprocessing v1 with Anomaly Detection - eXtreme Gradient Boosted Trees Regressor
\item BP 38: Average Blender
\item BP 39: Tree-based Algorithm Preprocessing v1 - eXtreme Gradient Boosted Trees Regressor with Early Stopping - Forest (10x)
\item BP 40: Tree-based Algorithm Preprocessing v1 - eXtreme Gradient Boosted Trees Regressor (learning rate =0.01)
\item BP 41: Gradient Boosting Model Preprocessing v4 - eXtreme Gradient Boosted Trees Regressor (learning rate =0.01)
\item BP 42: Tree-based Algorithm Preprocessing v15 - eXtreme Gradient Boosted Trees Regressor (learning rate =0.01)
\item BP 43: Regularized Linear Model Preprocessing v5 - Nystroem Kernel SVM Regressor
\item BP 44: Regularized Linear Model Preprocessing v5 - Support Vector Regressor (Radial Kernel)
\item BP 45: Missing Values Imputed - Smooth Ridit Transform - Partial Principal Components Analysis - Gaussian Process Regressor with Radial Basis Function Kernel
\item BP 46: Missing Values Imputed - Smooth Ridit Transform - Gaussian Process Regressor with Radial Basis Function Kernel
\item BP 47: Regularized Linear Model Preprocessing v5 - Auto-tuned K-Nearest Neighbors Regressor (Euclidean Distance)
\item BP 48: Missing Values Imputed - Smooth Ridit Transform - Gaussian Process Regressor with Mat\'ern Kernel (nu=1.5)
\item BP 49: Missing Values Imputed - Smooth Ridit Transform - Partial Principal Components Analysis - Gaussian Process Regressor with Mat\'ern Kernel (nu=1.5)
\item BP 50: Numeric Data Cleansing - Smooth Ridit Transform - Keras Slim Residual Neural Network Regressor using Adaptive Training Schedule (1 Layer: 64 Units)
\item BP 51: Missing Values Imputed - Decision Tree Regressor
\item BP 52: Numeric Data Cleansing - Ridge Regressor with Binned numeric features
\item BP 53: Tree-based Algorithm Preprocessing v2 - RandomForest Regressor
\item BP 54: Tree-based Algorithm Preprocessing v2 - RandomForest Regressor (Shallow)
\item BP 55: Tree-based Algorithm Preprocessing v1 - RandomForest Regressor
\item BP 56: Missing Values Imputed - RandomForest Regressor
\item BP 57: Missing Values Imputed - Smooth Ridit Transform - Gaussian Process Regressor with Mat\'ern Kernel (nu=2.5)
\item BP 58: Missing Values Imputed - Smooth Ridit Transform - Partial Principal Components Analysis - Gaussian Process Regressor with Mat\'ern Kernel (nu=2.5)
\item BP 59: Regularized Linear Model Preprocessing v2 - Ridge Regressor
\item BP 60 : Regularized Linear Model Preprocessing v12 - Ridge Regressor
\item BP 61: Regularized Linear Model Preprocessing v12 - Lasso Regressor
\item BP 62: Constant Splines - Ridge Regressor
\item BP 63: Numeric Data Cleansing - Smooth Ridit Transform - Keras Slim Residual Neural Network Regressor using Training Schedule (1 Layer: 64 Units)
\item BP 64: Tree-based Algorithm Preprocessing v1 - Light Gradient Boosted Trees Regressor with Early Stopping
\item BP 65: Tree-based Algorithm Preprocessing v1 - Dropout Additive Regression Trees Regressor  (15 leaves)
\item BP 66: Tree-based Algorithm Preprocessing v17 - ExtraTrees Regressor
\item BP 67: Missing Values Imputed - Text fit on Residuals (L2 /  Least-Squares Loss) - Generalized Additive Model
\item BP 68: Tree-based Algorithm Preprocessing v2 - LightGBM Random Forest Regressor
\item BP 69: Missing Values Imputed - RuleFit Regressor
\item BP 70: Missing Values Imputed - Text fit on Residuals (L2 /  Least-Squares Loss) - Generalized Additive2 Model
\item BP 71: Mean Response Regressor
\item BP 72: Missing Values Imputed - Eureqa Generalized Additive Model (40 Generations)
\item BP 73: Regularized Linear Model Preprocessing v5 - Support Vector Regressor (Linear Kernel)
\item BP 74: Missing Values Imputed - Eureqa Generalized Additive Model (1000 Generations)
\item BP 75: Missing Values Imputed - Eureqa Generalized Additive Model (10000 Generations)
\item BP 76: Numeric Data Cleansing - Smooth Ridit Transform - Binning of numerical variables - Keras Slim Residual Neural Network Regressor using Adaptive Training Schedule (1 Layer: 64 Units)
\item BP 77: Numeric Data Cleansing - Smooth Ridit Transform - One-Hot Encoding - Keras Deep Residual Neural Network Regressor using Training Schedule (2 Layers: 512, 512 Units)
\item BP 78: Regularized Linear Model Preprocessing v4 - TensorFlow Neural Network Regressor
\item BP 79: Neural Network Algorithm Preprocessing v4 - Keras Residual AutoInt Regressor using Training Schedule (3 Attention Layers with 2 Heads, 2 Layers: 100, 100 Units)
\item BP 80: Numeric Data Cleansing - Smooth Ridit Transform - Keras Deep Self-Normalizing Residual Neural Network Regressor using Training Schedule (3 Layers: 256, 128, 64 Units)
\item BP 81: Numeric Data Cleansing - Smooth Ridit Transform - Keras Deep Residual Neural Network Regressor using Training Schedule (3 Layers: 512, 64, 64 Units)
\item BP 82: Numeric Data Cleansing - Smooth Ridit Transform - One-Hot Encoding - Keras Wide Residual Neural Network Regressor using Training Schedule (1 Layer: 1536 Units)
\item BP 83: Neural Network Algorithm Preprocessing v4 - Keras Residual Cross Network Regressor using Training Schedule (3 Cross Layers, 4 Layers: 100, 100, 100, 100 Units)
\item BP 84: Neural Network Algorithm Preprocessing v1 - TensorFlow Deep Learning Regressor
\item BP 85: Neural Network Algorithm Preprocessing v3 - TensorFlow Deep Learning Regressor
\item BP 86: Neural Network Algorithm Preprocessing v5 - TensorFlow Deep Learning Regressor
\item BP 87: Neural Network Algorithm Preprocessing v4 - Keras Residual Neural Factorization Machine Regressor using Training Schedule (2 Layers: 100, 100 Units)
\end{itemize}

\section{Building the reduced basis and empirical interpolant} \label{app:rb-eim}

The reduced basis is built using the Reduced Basis algorithm~(\cite{prud'homme:70}, see also~\cite{Field:2011mf} and citations therein for early applications in waveform modeling), being implemented in Python. The training set is composed of $50$ waveforms from the fiducial model, equally spaced in the mass ratio $q\in[1, 10]$. After building the reduced basis we construct the empirical interpolant using the EIM algorithm~\cite{Barrault2004667, Maday_2009, sorensen2010}.

We assess the representation accuracy of the RB and EIM approaches in the $L_2$ norm. The RB error for a waveform $h$ is defined as
\be
{\cal E } := {\cal E }(h) =  \| h -{\cal P}_n h   \|^2 \, ,  \label{eq:rep-rb}
\ee
where ${\cal P}_n$ is the orthogonal projection onto the basis of size $n$. The basis is hierarchically built until reaching a tolerance $\epsilon$ for the representation error of any waveform in the training set:
\be
{\cal E }(h) \leq \epsilon \quad \forall h \in \text{TS}_\text{rb} \, .  \label{eq:tol-rb}
\ee

Analogously, the interpolation error is defined as
\be
{\cal E}' := {\cal E}' (h) =  \| h -\cI_n [h]   \|^2 \, ,  \label{eq:rep-eim}
\ee

where $\cI_n$ is the EIM interpolant associated to the basis.

\begin{figure}[h!]
    \centering
    \includegraphics[width=0.48\textwidth]{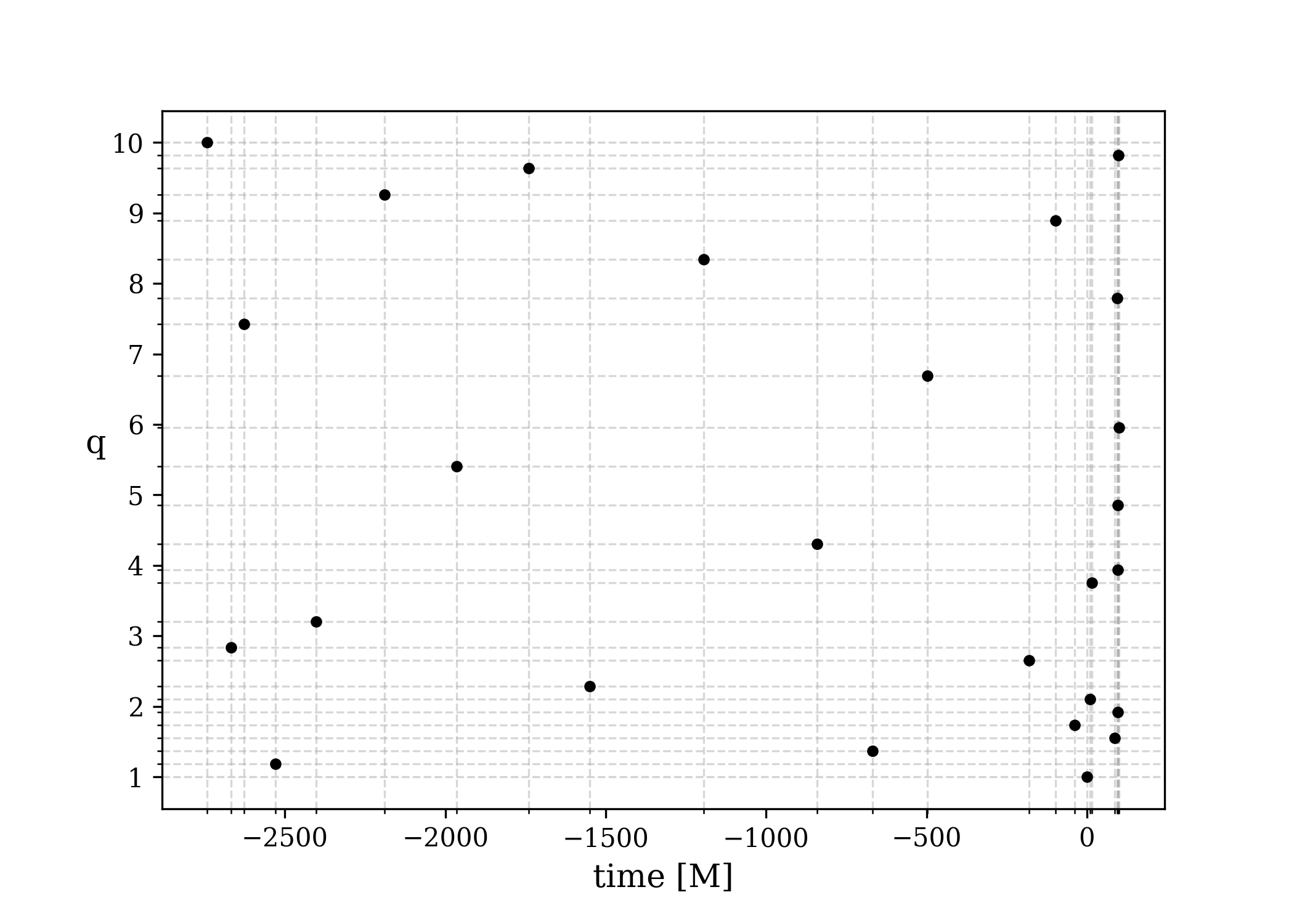}
    \includegraphics[width=0.48\textwidth]{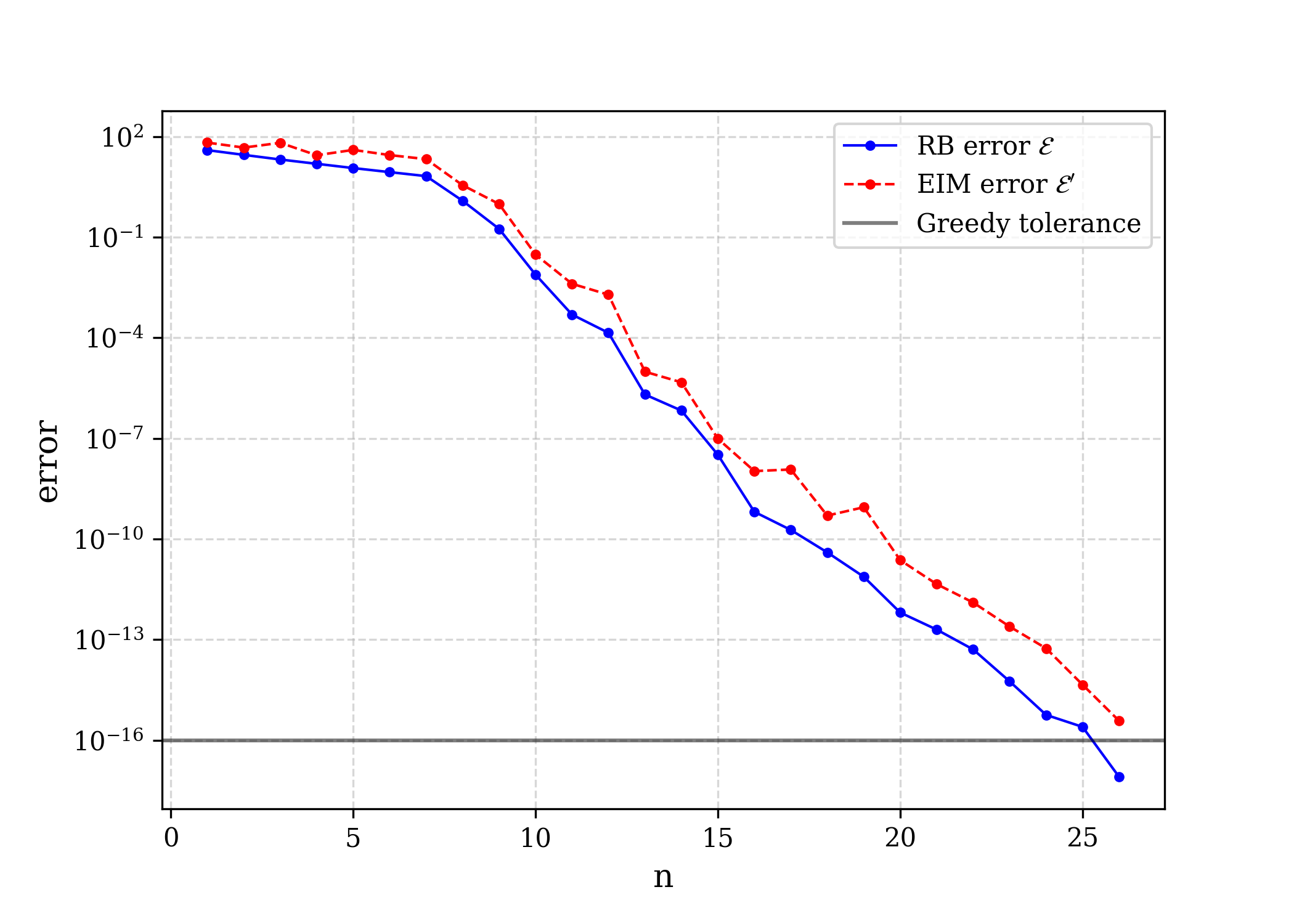}
    \caption{{\bf Left}: Illustration of the $26$ greedy points ($q$ axis) vs the $26$ EIM nodes (time axis) chosen by the RB/EIM algorithms. {\bf Right}: Representation errors in the training set \TSrb{} for both algorithms as function of the reduced basis dimensionality $n$. For $n=26$ they reach machine precision.}
    \label{fig:errs-gp}
\end{figure}

As described in~\cref{sec:intro,sec:feat-eng}, both algorithms comprise information related to the waveforms' parameter and time variations. This is done by selecting a suitable and minimal set of greedy parameter points and EIM time nodes which later serve to build the RB basis and the EIM interpolant. In Figure~\ref{fig:errs-gp} we show (left) the greedy selection of both algorithms along (right) the representation errors achieved by each one as a function of the dimensionality $n$ of the basis.

\begin{figure}[h!]
    \centering
    \includegraphics[width=0.7\textwidth]{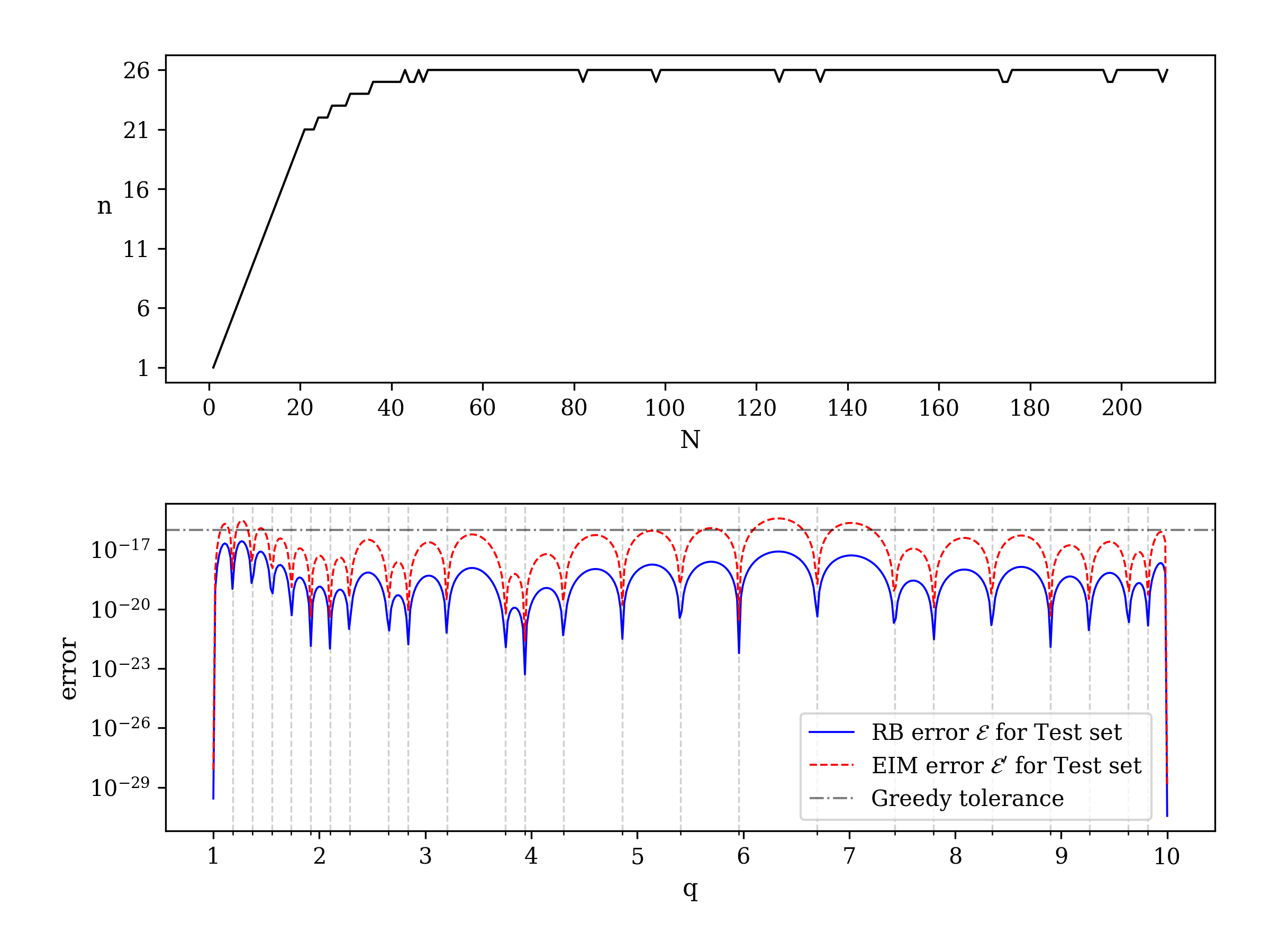}
    \caption{{\bf Top}: RB dimensionality $n$ vs. training number $N$. We observe that the RB dimensionality converges to $26$ for $N \sim 50$, which is the number of training waveforms used to build the basis for this work. {\bf Bottom}: Test curves to assess the representation accuracy of both approaches, RB and EIM, for a set of $500$ test waveforms equally spaced in the mass-ratio $q$. Vertical dashed lines correspond to greedy points. These are sectors of exact representation for the reduced basis, which is reflected in the look-down error peaks for the solid curve. Note that this behavior is also displayed by the interpolant.}
    \label{fig:convergence-test}
\end{figure}

We make some remarks.
\begin{itemize}
    \item In order to achieve machine precision $26$ reduced basis waveforms are needed. To that end we chose a greedy tolerance $\epsilon = 10^{-16}$.
    \item Greedy and EIM points distributions concentrate in the low-mass-ratio and merger-ringdown sectors (left panel of Figure~\ref{fig:errs-gp}). These correspond to sectors in which waveforms change more rapidly than the rest of the parameter-time domain.
    \item In the right panel of Figure~\ref{fig:errs-gp}, ${\cal E} \leq {\cal E}'$. This is a general fact, since both, orthogonal projection $\cP_n h$ and interpolation $\cI_n[h]$, are linear combinations of reduced basis elements. It is well known that the best linear combination in the least-squares sense is the first one, $\cP_n h$. Any other representation in the form of linear expansions of the basis will perform worse in the $L_2$ norm than the optimal one ($\cP_n h$); this includes interpolation.
    \item We found that the reduced basis dimensionality $n$ converges to $26$ for large training sets (top panel of Figure~\ref{fig:convergence-test}), meaning that a reduced basis with $n=26$ captures all the structure related to waveforms in the range $q\in[1, 10]$ at a tolerance $\epsilon = 10^{-16}$.
    \item Although one expects some quality loss when moving from a RB representation to EIM interpolation, this loss is negligible when $n=26$. In the bottom panel of Figure~\ref{fig:convergence-test} we assess the representation capacity of both, RB and EIM approaches, in a test set composed of $500$ waveforms equally spaced in the mass ratio $q\in[1, 10]$. As shown, all interpolation errors fall below or close to the roundoff line $\epsilon = 10^{-16}$.

\end{itemize}

\section{Main regression approaches} \label{sec:regression}

\noindent{\bf Gaussian Process Regression}.-- GPR uses Bayesian statistics to infer models from data~\cite{gpr2006Rasmussen}. Its ultimate goal is to build suitable probability distributions over the space of possible models by performing Bayesian inference on observed data:
$$
p(f | {\mathcal D}) = \frac{p({\mathcal D} | f) p(f)}{p({\mathcal D})}\,.
$$
Before any data is seen, the prior $p(f)$ describes an initial guess distribution over possible models $f$. When observed data $\mathcal D$ is available, the prior gets actualized and gives a posterior distribution $p(f | {\mathcal D})$ over models which makes better estimates for regression and the corresponding uncertainties.

Strictly speaking, a Gaussian process is a set of random variables for which any finite subset has a joint Gaussian distribution. In this context, a model $f$ can be seen as a infinite-dimensional stochastic process from which any finite subsample is jointly Gaussian. This kind of processes is completely defined by a mean function $\mu(x)$ and a covariance/{\it kernel} function $k(x, x')$. The kernel function plays a capital role in Bayesian regression since it determines the kind of functions that will be available for modeling and related properties such as degree of smoothness, characteristic lengths, etc. The specification of the kernel determines the type of GPR to carry out. A prototypical example is the Radial Basis Function (RBF) or Gaussian kernel defined by
$$
k_{RBF}(x, x') := \exp( - \frac{1}{2l^2} | x - x' |^2)\,.
$$
This kernel is the most common and widely used in the field. Its name comes from the fact that the $(x, x')$ dependency enters the function only through its normed difference $| x -x' |$, a feature also present in SVM approaches. The parameter $l$ is the characteristic length-scale and controls the scale of variation of inferred models. More general kernels can be built from the RBF by the addition of parameters to the function. Once a kernel is defined, all free parameters can be optimized for better performance. In fact, GPR enters the AutoML flow through hyperparameter optimization of these variables.

Another important family of kernels is the Mat\'ern class. It is described by covariance functions of the form
$$
k_{M}(r) := \frac{ 2 ^{1- \nu} } {\Gamma(\nu)} \Bigg( \frac{\sqrt{2\nu} r} { l } \Bigg)^\nu K_\nu \Bigg(  \frac{\sqrt{2\nu} r} { l } \Bigg)\,
$$ 
where $r := |x - x'|$, $\nu, l$ are positive parameters and $K_\nu$ is a modified Bessel function. In some cases Mat\'ern kernels are preferred over RBF ones since the latter carries strong smoothness assumptions that may turn out to be unrealistic in many physical scenarios~\cite{gpr2006Rasmussen}.

Besides the advantage of GPR approaches at providing not only predictive models but an uncertainty estimation for them, its algorithm scales as $\bigo{n^3}$ with $n$ the size of the dataset, something that may cause computational bottlenecks for large data.

\noindent{\bf Eureqa}.-- Eureqa models implement symbolic regression using evolutionary algorithms through genetic programming~\cite{Koza:1992:GPP:138936, Schmidt03042009, Schmidt2010}. As opposed to more standard approaches such as polynomial regression, the goal of symbolic regression is to find not only the best set of parameters for fitting analytical forms to data but the form itself. This is done first by specifying a dictionary of building blocks, e.g. elementary functions, that serves to build more complex forms. Naturally, this leads to a problem of combinatorial complexity since the search space is composed of all the possible combinations of building blocks with their respective free parameters. To overcome this, an evolutionary search, in mimicking the ways of nature, is performed across many possible combinations, relaxing the search complexity by introducing randomness in the selection process. Through the evolution of populations using e.g. random mutations or crossover, different models compete with each other following some optimization criteria until the best individuals are left. In Eureqa, the best models are ranked following a compromise between accuracy and complexity  summarized in a Pareto front curve. Early applications of these methods in gravitational wave modeling can be found in \cite{tiglio2021ab}.

\noindent{\bf Support Vector Machines}.-- SVM algorithms~\cite{Boser1996,cortes1995support} approximate functions by fitting expansions of the form
$$
f(x) = \sum_{i=1}^n \alpha_i k(x, x_i) \, , 
$$
where $\alpha_i$ are free parameters, $x_i$ are data points, and $k(x, x_i)$ is a kernel function in the same sense as in GPR. The kernel plays the role of a generalized inner product between nonlinear feature maps. These maps take the input vectors $x,\,x_i$ and project them to other feature spaces. In those one can perform trivial linear regression.

SVM is computationally efficient since convex optimization can be directly applied as the problem is linear in the free parameters $\alpha_i$. It has the obvious advantage over linear approaches of being able to model nonlinearities present in training data, something that direct linear regression can get to perform very poorly. As in GPR, kernel functions play a prominent role. For example, kernel functions may be computationally more tractable than using feature maps in some cases. Even worse, feature maps can be infinite dimensional and therefore impossible of computer implementation.

\bibliography{references}

\end{document}